# Semi-empirical pressure-volume-temperature equation of state; MgSiO$_3$ perovskite is an example

József Garai

*Department of Earth Sciences, Florida International University, Miami, FL 33199, USA*

**Abstract**

Simple general formula describing the pressure-volume-temperature relationships (p-V-T) of elastic solids is constructed from theoretical considerations. The semi-empirical equation of state (EoS) was tested to experiments of perovskite 0-109 GPa and 293-2000 K. The parameters providing the best fit are: $B_o = 267.5$ GPa, $V_o = 24.284$ cm$^3$, $\alpha_o = 2.079 \times 10^{-5}$ K$^{-1}$, $\frac{\partial B_o}{\partial p} = 1.556$ and $\frac{\partial \alpha_o}{\partial p} = -1.098 \times 10^{-7}$ K$^{-1}$GPa$^{-1}$. The root-mean-square-deviations (RMSD) of the residuals are 0.043 cm$^3$, 0.79 GPa, and 125 K for the molar volume, pressure, and temperature respectively. These RMSD values are in the range of the uncertainty of the experiments, indicating that the five parameters semi-empirical EoS correctly describes the p-V-T relationships of perovskite.

Separating the experiments into 200 K ranges the semi-empirical EoS was compared to the most widely used finite strain, interatomic potential, and empirical isothermal EoSs such as the Birch-Murnaghan, the Vinet, and the Roy-Roy respectively. Correlation coefficients, RMSD's of the residuals and Akaike Information Criteria were used for evaluating the fitting. Based on these fitting parameters under pure isothermal conditions the semi-empirical p-V EoS is slightly weaker than the Birch-Murnaghan and Vinet EoSs; however, the semi-empirical p-V-T EoS is superior in every temperature range to all of the investigated conventional isothermal EoSs.

## I. INTRODUCTION

The relationships among the pressure, the volume, and the temperature are described by the EoS. The basic relationship between the volume and the temperature is described by the definition of the volume coefficient of expansion [$\alpha_V$]

$$\alpha_{V_p} \equiv \frac{1}{V}\left(\frac{\partial V}{\partial T}\right)_p \tag{1}$$

The relationship between the pressure and the volume is given by the isothermal bulk modulus [$B_T$]

$$B_T \equiv -V\left(\frac{\partial p}{\partial V}\right)_T . \tag{2}$$

For the validity of equation (2) it is assumed that the solid is homogeneous, isotropic, non-viscous and has linear elasticity. It is also assumed that the stresses are isotropic; therefore, the principal stresses can be identified as the pressure $p = \sigma_1 = \sigma_2 = \sigma_3$.

Experiments show that both the volume coefficient of expansion and the isothermal bulk modulus are pressure and temperature dependent; therefore, it is necessary to know the derivatives of these parameters.

$$\left(\frac{\partial \alpha_V}{\partial T}\right)_p ; \left(\frac{\partial \alpha_V}{\partial p}\right)_T ; \left(\frac{\partial B_T}{\partial T}\right)_p ; \left(\frac{\partial B_T}{\partial p}\right)_T \tag{3}$$

The schematic relationships between the thermodynamic quantities and parameters are shown on Fig. 1-a. A universal EoS must cover the entire pressure and temperature range; therefore, it is necessary to incorporate all of the derivatives of the volume coefficient of expansion and the isothermal bulk modulus. The EoS is an integral part of the Helmholtz and Gibbs free energies and an important tool describing the pressure-temperature-composition relationships[1]. There is no single expression known for universal (p-V-T) EoS[2,3].

## II. DESCRIBING THE P-V-T RELATIONSHIP

In order to overcome the complexity of the EoS, the common practice is that the temperature of the substance is raised first and then the substance is compressed along the isotherm of interest[4,5]. The relevant equations are called the thermal and the isothermal EoS respectively. The thermal EoS is used to calculate the volume at atmospheric pressure and temperature T $[V_{0,T}]$. It is also necessary to know the temperature affect on the bulk modulus $[B_0(T)]$. Using the values of the volume and the bulk modulus at the corresponding temperature the isothermal EoS calculates the affect of pressure by incorporating the first and the second derivates of the bulk modulus, $\left(\frac{\partial B}{\partial p}\right)_T$ and $\left(\frac{\partial^2 B}{\partial p^2}\right)_T$ at the given temperature.

### A. Thermal EoS

The simplest thermal equation of state is derived by the integration of the thermodynamic definition of the volume coefficient of thermal expansion Eq.(1)

$$V_0(T) = V_0(T_0) e^{\int_{T_0}^{T} \alpha_{V_p}(T) dT} \qquad (4)$$

If a wider temperature range is considered then the temperature dependence of the volume coefficient of thermal expansion should be known. Knowing the first derivative of this parameter allows one to calculate the high temperature values:

$$\alpha_{V_p}(T) = \alpha_{V_p}(T_0) + \left(\frac{\partial \alpha_V}{\partial T}\right)_P (T - T_0). \qquad (5)$$

The thermodynamic Gruneisen-Anderson parameter $[\delta_T]$ is defined as[6]:

$$\delta_T = \left(\frac{\partial \ln B_T}{\partial \ln \rho}\right)_P = -\frac{1}{\alpha_V}\left(\frac{\partial \ln B_T}{\partial T}\right)_P \qquad (6)$$

Assuming that the solid at higher temperatures follows classical behavior, then the product of $\alpha_{V_p} B_T$ is constant and the Gruneisen-Anderson parameter is independent of temperature, Anderson et al.[7] and Shanker[8] proposed the following isobaric EoS:

$$V = V_0 \left[1 - \alpha_{V_0} \delta_0 (T - T_0)\right]^{\frac{1}{\delta_0}} \qquad (7)$$

where the subscript zero values of the parameters refers to the initial temperature of $T_0$.

Assuming that the product of $\alpha_{V_p} B_T$ is constant and the Gruneisen-Anderson parameter changes linearly with the volume, the following EoS has been proposed by Kumar[9] and Kushwah et al.[10]

$$V = V_0 \left\{1 - \frac{1}{A} \ln\left[1 - \alpha_{V_0} A (T - T_0)\right]\right\} \qquad (8)$$

where $A = \delta_0 + 1$.

Thermal EoS have been suggested by Akaogi and Navrotsky[11,12], assuming that the thermal expansion is quadratic in the temperature, and independent of pressure

$$V_p = V_0 \left[1 + \alpha_{V_0}(T - T_0) + \alpha'_{V_0}(T - T_0)^2\right], \qquad (9)$$

where $\alpha'_{V_0}$ is the temperature derivative of $\alpha_V$ at temperature $T = T_0$. Taking into consideration the affect of the pressure the equation can be written as:

$$V_{p,T} = V_0 \left[1 + \frac{B'_0 (p - p_0)}{B_0}\right]^{\frac{-1}{K'_0}} \left[1 + \alpha_{V_0}(T - T_0) + \alpha'_{V_0}(T - T_0)^2\right], \qquad (10)$$

Fei and Saxena[13] revised the quadratic relationship of Eq. (10) and proposed the following empirical expression:

$$V_p = V_0 \left[1 + \alpha_{V_0}(T - T_0) + \frac{1}{2}\alpha'_{V_0}(T - T_0)^2 - \alpha_{V_0}(T - T_0)^{-1}\right]. \qquad (11)$$

Assuming linear change as a function of temperature in the volume coefficient of thermal expansion leads to the following expression[14]

$$V_p = V_0 e^{\alpha(T-T_0) + \frac{1}{2}\alpha'(T-T_0)^2} \qquad (12)$$

The proposed general expression of Eq. (12) for bcc iron is:

$$V_p = V_0 \left\{ 1 + B_0' \left[ B_0 - \left(\frac{\partial B}{\partial T}\right)_0 (T-T_0) \right] \frac{p-p_0}{B_0^2} \right\}^{\frac{-1}{K_0'}} e^{\alpha (T-T_0) + \frac{1}{2}\alpha'(T-T_0)^2} \quad (13)$$

An empirical expression has been given by Plymate and Stout[15]

$$V = V_0 \left\{ \left[1 + \left(\frac{\partial B_T}{\partial T}\right)_0 \left(\frac{T-T_0}{B_0}\right)\right]^{-\frac{1}{B_0'}} e^{\left[\alpha v_0 + \left(\frac{\partial B_T}{\partial T}\right)_0 \frac{1}{B_0 B_0'}\right](T-T_0) + \left[\alpha' v_0 + \left(\frac{\partial B_T}{\partial T}\right)_0^2 \frac{1}{B_0^2 B_0'}\right]\frac{(T-T_0)^2}{2}} \right\}. \quad (14)$$

**B. The temperature effect on the bulk modulus**

The effect of the temperature on the bulk modulus is discussed in details in ref. 16. Assuming constant value for the product of the volume coefficient thermal expansion and the bulk modulus allows deriving an analytical solution for the temperature dependence of the bulk modulus at temperatures higher than the Debye temperature[16].

$$B_T = e^{-\int_{T=0}^{T} \delta_T \alpha_V dT} B_{T=0}. \quad (15)$$

where $\delta_T$ is the isothermal Anderson-Grüneisen parameter given by:

$$\delta_T = -\frac{1}{\alpha_{V_p} B_T} \left(\frac{\partial B_T}{\partial T}\right)_P. \quad (16)$$

**C. Isothermal EoS**

The determined values of the volume and the bulk modulus at temperature T can be used as initial parameters for an isothermal EoS. The isothermal equations of states follow finite strain, interatomic potential, or empirical approach.

*1. Finite-strain EoS*

The Birch-Murnaghan EoS[17-19] assumes that the strain energy of a solid can be expressed as a Taylor series in the finite Eulerian strain, $f_E$. Expansion to fourth order in the strain yields an EoS:

$$p = 3B_0 f_E (1+2f_E)^{\frac{5}{2}} \left\{ 1 + \frac{3}{2}(B'-4)f_E + \frac{3}{2}\left[ B_0 B'' + (B'-4)(B'-3) + \frac{35}{9} \right] f_E^2 \right\} \quad (17)$$

where $f_E$ is

$$f_E = \frac{\left(\frac{V_0}{V}\right)^{\frac{2}{3}} - 1}{2}. \quad (18)$$

The most widely used isothermal EoS is the third-order Birch-Murnaghan. Quite recently Sushil et al.[20] used n=1 instead of the n=2 in the Eulerian strain measure

$$f_E = \frac{\left(\frac{V_0}{V}\right)^{\frac{n}{3}} - 1}{n} = \left(\frac{V_0}{V}\right)^{\frac{1}{3}} - 1 \quad (19)$$

and using the method of Stacey[21] proposed a modified three-parameter Eulerian strain EoS,

$$p = \frac{9}{2} B_0 \left( -A_1 x^{-\frac{4}{3}} + A_2 x^{-\frac{5}{3}} - A_3 x^{-2} + A_4 x^{-\frac{7}{3}} \right). \quad (20)$$

where

$$x = \frac{V}{V_0}, \quad A_1 = B_0 B_0'' + (B_0' - 3)^2 + \frac{26}{9}$$

$$A_2 = 3B_0 B_0'' + (B_0' - 3)(3B_0' - 8) + \frac{66}{9}$$

$$A_3 = 3B_0 B_0'' + (B_0' - 3)(3B_0' - 7) + \frac{60}{9}$$

and $\quad A_4 = B_0 B_0'' + (B_0' - 3)(B_0' - 2) + \frac{20}{9}$

The authors claimed that their modified Eulerian strain EoS is more rapidly convergent than the Birch-Murnaghan EoS.

2 Inter-atomic potential EoSs

The theoretical base for the interatomic potential EoS lays in the thermodynamic relationship

$$p = T\left[\frac{\partial p}{\partial T}\right]_V - \left[\frac{\partial U}{\partial V}\right]_{T,m} \quad (21)$$

where m stands for a mol quantity. Neglecting the thermal pressure[22,23] and approaching the second term, the so-called internal pressure, with the volume derivative of the biding energy allows determining the pressure-volume relationship. The resulting EoS contains three parameters, the zero pressure values of the molar volume, the isothermal bulk modulus, and the pressure derivative of the bulk modulus.

Using the potential function proposed by Mie and extended by Grunesisen[24,25]

$$U(r) = -\frac{A}{r^m} - \frac{B}{r^n} = -\frac{A}{V^{\left(\frac{m}{3}\right)}} + \frac{B}{V^{\left(\frac{n}{3}\right)}}. \tag{22}$$

where r is the interatomic spacing and A, B, m, and n are constants (not necessarily integers) the P-V equation of state can be written as[26]:

$$p = \frac{3B_T(0)}{m-n}\left[\left(\frac{V_o}{V}\right)^{\left(\frac{m+3}{3}\right)} - \left(\frac{V_o}{V}\right)^{\left(\frac{n+3}{3}\right)}\right]. \tag{23}$$

The so-called universal EoS derived by Rose from a general inter-atomic potential[27], which was promoted by Vinet[22,23] is also commonly used:

$$p = 3B_0 \frac{1-f_V}{f_V^2} e^{\left[\frac{3}{2}(B'-1)(1-f_V)\right]} \tag{24}$$

where

$$f_V = \left(\frac{V}{V_0}\right)^{\frac{1}{3}}. \tag{25}$$

The Vinet EoS gives very accurate results for simple solids at very high pressure.

Some authors[28,29] pointed out that there is a restriction on Eq. (24) when it is applied to high-pressure phase solids under low pressure conditions. The use of $p = 0$ and $V = V_0$ is sometimes arbitrary since the high pressure phase might not exist under this condition. In order to overcome on this problem Fang[30] suggested modifying the original Vinet equation (24) by introducing an additional parameter. In this modified equation it was assumed that the isothermal bulk modulus varies linearly with the pressure.

Precise knowledge of the interatomic forces in the stress-free state and their variation with pressure and temperature would allow calculating all the thermodynamic properties. The lack of such knowledge has resulted in many two and three-parameters empirical EoSs.

*3 Empirical EoSs*

Empirical EoSs can be divided into two major groups. One uses the original Eulerian strain or Interatomic potential EoSs and refines their parameters in order to find a better fit to experiments[31-34]. The other approach is to find a mathematical function which gives the best fit to the experiments[35-38].

Roy and Roy[39] give a good review and evaluate the fittings of the currently used EoSs. Their proposed[40] three parameter empirical EoS is

$$V = V_0\left[1 - \frac{\ln(1+ap)}{b+cp}\right], \qquad (26)$$

where

$$a = \frac{1}{8B_0}\left[3(B_0'+1)+(25B_0'^2+18B_0'-32B_0B_0''-7)^{\frac{1}{2}}\right]$$

$$b = \frac{1}{8}\left[3(B_0'+1)+(25B_0'^2+18B_0'-32B_0B_0''-7)^{\frac{1}{2}}\right]$$

$$c = \frac{1}{16}\left[3(B_0'+1)+(25B_0'^2+18B_0'-32B_0B_0''-7)^{\frac{1}{2}}\right](B_0'+1)-\frac{1}{8}\left[3(B_0'+1)+(25B_0'^2+18B_0'-32B_0B_0''-7)^{\frac{1}{2}}\right]$$

They used shock compression data of different metals[41] and the calculated EoS of halite[42] to evaluate the proposed equation up to ultra high pressures.

The empirical nature of these equations usually leads to a lack of generality and careful inspection reveals that a particular equation is typically gives excellent fitting only for special substances or a specially selected pressure and/or temperature range.

Many of the parameters in the EoS are inter-related, which adds to the complexity of calculations. The optimum values of each of the interrelated parameters have to be determined by confidence ellipses[5,43]. The thermodynamic description of solids is complicated, time consuming, labor intensive, and expensive.

### III. FUNDAMENTAL COMPONENTS OF THE VOLUME IN SOLID PHASE

Avogadro's principle does not apply to solids contrarily to gasses. Matter in solid phase occupies an initial volume $[V_o]$ at zero pressure and temperature.

$$V_o = nV_o^m, \tag{27}$$

where n is the number of moles and $V_o^m$ is the molar volume of the substance at zero pressure and temperature. The pressure modifies this initial volume by inducing elastic deformation while the temperature by causing thermal deformation. Using equations (1) and (2) the actual volume at given pressure and temperature can be calculated[44] by allowing one of the variables to change while the other one held constant

$$[V_T]_{p=0} = V_0 e^{\int_{T=0}^{T} \alpha V_{p=0} dT} \quad \text{or} \quad [V_p]_{T=0} = V_0 e^{-\int_{p=0}^{p} \frac{1}{B_{T=0}} dp} \tag{28}$$

and then

$$[V_T]_p = [V_p]_{T=0} e^{\int_{T=0}^{T} \alpha V_p dT} \quad \text{or} \quad [V_p]_T = [V_T]_{p=0} e^{-\int_{p=0}^{p} \frac{1}{B_T} dp} \tag{29}$$

These two steps might be combined into one and the volume at a given p, and T can be calculated:

$$V_{p,T} = V_0 e^{\int_{T=0}^{T} \alpha V_{p=0} dT - \int_{p=0}^{p} \frac{1}{B_T} dp} = V_0 e^{\int_{T=0}^{T} \alpha V_p dT - \int_{p=0}^{p} \frac{1}{B_{T=0}} dp} \tag{30}$$

The total volume change related to the temperature will be called thermal volume $[V^{th}]$ while the total volume change resulted from elastic deformation will be called elastic volume $[V^{el}]$. The thermal volume at zero pressure is

$$[V_T^{th}]_{p=0} = V_0 \left( e^{\int_{T=0}^{T} \alpha_V dT} - 1 \right), \tag{31}$$

while the elastic volume at zero temperature is

$$[V_p^{el}]_{T=0} = V_o \left( e^{-\int_{p=0}^{p} \frac{1}{B_T} dp} - 1 \right).  \quad (32)$$

The thermal volume at pressure p is

$$[V_T^{th}]_p = [V_T^{th}]_{p=0} e^{-\int_{p=0}^{p} \frac{1}{B_T} dp} = V_o e^{-\int_{p=0}^{p} \frac{1}{B_T} dp} \left( e^{\int_{T=0}^{T} \alpha_V dT} - 1 \right),  \quad (33)$$

and the elastic volume at temperature T is

$$[V_p^{el}]_T = [V_p^{el}]_{T=0} e^{\int_{T=0}^{T} \alpha_V dT} = V_o e^{\int_{T=0}^{T} \alpha_V dT} \left( e^{-\int_{p=0}^{p} \frac{1}{B_T} dp} - 1 \right).  \quad (34)$$

The actual volume is the sum of the volume components:

$$[V_T]_p = V_o + [V_p^{el}]_{T=0} + [V_T^{th}]_p \quad \text{or} \quad [V_p]_T = V_o + [V_p^{el}]_T + [V_T^{th}]_{p=0}  \quad (35)$$

Since

$$[V_T]_p = [V_p]_T  \quad (36)$$

from Eq (35) follows that

$$[V_T^{th}]_p - [V_T^{th}]_{p=0} = [V_p^{el}]_T - [V_p^{el}]_{T=0}.  \quad (37)$$

The compressed part of the thermal volume is the same as the expanded part of the elastic volume. Since the volume difference in Eq. (37) both temperature and pressure dependent I will call this volume difference to thermo-elastic volume $[\Delta V_p^{th-el}]_T$

$$[V_p^{th-el}]_T = [V_T^{th}]_p - [V_T^{th}]_{p=0} = [V_p^{el}]_T - [V_p^{el}]_{T=0}.  \quad (38)$$

The thermoelastic volume can be calculated as:

$$[V_p^{th-el}]_T = V_o \left( e^{\int_{T=0}^{T} \alpha_V dT} - 1 \right) \left( e^{-\int_{p=0}^{p} \frac{1}{B_T} dp} - 1 \right).  \quad (39)$$

It can be concluded that the actual volume comprises from four distinct volume parts, initial volume, thermal volume at zero pressure $[V_T^{th}]_{p=0}$, elastic volume at zero temperature $[V_p^{el}]_{T=0}$, and thermo-elastic volume $[V_p^{th-el}]_T$ (Fig. 2).

$$V = V_o + [V_T^{th}]_{p=0} + [V_p^{el}]_{T=0} + [V_p^{th-el}]_T. \tag{40}$$

These fundamental volume components are related to the thermo-physical variables as:

$$V_o = f(n)_{T=0; p=0}, \tag{41}$$

$$[V_T^{th}]_{p=0} = f(T)_{n; p=0}, \tag{42}$$

$$[V_p^{el}]_{T=0} = f(p)_{n; T=0}, \tag{43}$$

and

$$[V_p^{th-el}]_T = f(T, p)_n. \tag{44}$$

Eqs. (1) and (2) give the right value for the volume because the integrations of

$$\int \frac{1}{x} dx = \ln(x) \tag{45}$$

and

$$\int \frac{1}{a+x} dx = \ln(a+x). \tag{46}$$

are numerically equivalent. However, the proper physical description of the relationships requires addressing the constants and the variables appropriately.

**IV. THE PROPOSED EoS**

The following four equations are suggested for describing the p-V-T relationships of elastic solids. The relationship between the initial volume and the number of moles is defined by Eq. (27) as:

$$V_o \equiv nV_o^m \qquad \Rightarrow \qquad V_o = f(n)_{T=0;\ p=0}. \tag{47}$$

The elastic volume and the pressure relationship can be defined as:

$$B_{T=0} \equiv -(V_o + V^{el})\left(\frac{\partial p}{\partial V^{el}}\right)_{T=0} \qquad \Rightarrow \qquad V_p^{el} = f(p)_{T=0;\ V_o}. \tag{48}$$

The relationship between the temperature and the thermal volume at a given pressure is given as:

$$\alpha_{V_p} \equiv \frac{1}{V_{p,T=0} + V_p^{th}}\left(\frac{\partial V^{th}}{\partial T}\right)_p \qquad \Rightarrow \qquad \left[V_T^{th}\right]_p = f(T)_{p;\ V_o}. \tag{49}$$

The sum of the fundamental volume components restores the actual volume as:

$$V_{p,T} \equiv V_o + V_p^{el} + \left[V_T^{th}\right]_p \qquad \Rightarrow \qquad V_{p,T} = f\left(V_o;\ V^{el};\ \left[V_T^{th}\right]_p\right). \tag{50}$$

The simplest EoS resulting from these four definitions [Eqs. (47)-(50)] is

$$V_{p,T} = nV_o^m e^{\int_{T=0}^{T}\alpha_{V_p}dT - \int_{p=0}^{p}\frac{1}{B_{T=0}}dp} = nV_o^m e^{\alpha_{V_p}T - \frac{p}{B_{T=0}}}. \tag{51}$$

By definition [Eq. (48)] the temperature derivative of the bulk modulus is zero. This zero value is also consistent with theory since the interatomic energies are independent of the temperature. Thus the temperature derivatives of this parameter should remain zero at any temperature

$$\frac{\partial B}{\partial T} = 0. \tag{52}$$

Assuming that the pressure dependence of the bulk modulus can be described by a linear [a] and a quadratic factor [b] results in

$$B = B_o + ap + bp^2, \tag{53}$$

where

$$B_o \equiv \lim_{p \to 0} B_{T=0}. \tag{54}$$

Using Eq. (12) and substituting $0.5\alpha'$ with a constant multiplier [d] the temperature dependence of the volume coefficient of thermal expansion at zero pressure can be described as:

$$\alpha_{V_{p=0}} = \alpha_o + dT. \quad (55)$$

where

$$\alpha_o \equiv \lim_{T \Rightarrow 0} \alpha_{V_{p=0}} \quad (56)$$

The thermal volume at a given pressure comprises from the thermal volume at zero pressure and the thermal elastic volume Eq. (38). In order to take into account the effect of pressure on the thermal volume a constant pressure derivative [c] is introduced

$$\alpha_{p,T=0} = \alpha_o + cp. \quad (57)$$

It is also assumed that the temperature factor [d] in Eq. (55) changes in the same rate as $\alpha_{p,T=0}$ as a function of pressure. Introducing a normalizing pressure factor from Eq. (57)

$$\left(1 + \frac{cp}{\alpha_o}\right) \quad (58)$$

gives the pressure and temperature dependence of the volume coefficient of thermal expansion

$$\alpha_{T,p} = \alpha_{p,T=0} + \left(1 + \frac{cp}{\alpha_o}\right)dT. \quad (59)$$

Substituting Eq. (57) into Eq. (59) results in

$$\alpha_{T,p} = \alpha_o + cp + \left(1 + \frac{cp}{\alpha_o}\right)dT = \left(1 + \frac{cp}{\alpha_o}\right)(\alpha_o + dT). \quad (60)$$

Combining Eqs. (51), (53) and (60) gives the p-V-T relationship for elastic solids as:

$$V = nV_o^m e^{\frac{-p}{ap+bp^2+B_o} + \left(1+\frac{cp}{\alpha_o}\right)(\alpha_o+dT)T} \quad \text{(S-E-7)} \quad (61)$$

Equation (61) contains seven parameters $V_o^m$; $K_o$; $\alpha_o$; a; b; c; and d; and will be labeled as Semi-empirical 7 (S-E-7). The thermodynamic relationships incorporated into Eq. (61) are shown on Fig 1-b. Equation (61) has an analytical solution for the temperature

$$T = \frac{-\alpha_o \pm \sqrt{\alpha_o^2 + 4d \dfrac{\ln\left(\dfrac{V}{V_o}\right) + \dfrac{p}{ap+bp^2+B_o}}{1+\dfrac{cp}{\alpha_o}}}}{2d} \quad \text{(S-E-7)}. \tag{62}$$

The pressure can be determined by repeated substitutions as:

$$p = \lim_{n \to \infty} f^n(p) \tag{63}$$

where

$$f^n(p) = \left(B_o + ap_{n-1} + bp_{n-1}^2\right)\left[\left(1 + \frac{cp_{n-1}}{\alpha_o}\right)(\alpha_o + dT)T - \ln\left(\frac{V}{V_o}\right)\right]; \quad n \in \mathbb{N}^* \tag{64}$$

and $p_0 = 0$.

The convergence of Equation (63)-(64) depends on the pressure. For the maximum pressure used in this study (up to 100 GPa) n = 10 gives sufficiently good result.

If the pressure derivative of the bulk modulus is constant and the temperature has no effect on the volume coefficient of thermal expansion that is b=0 and d=0 then Eqs. (61)- (64) can be simplified as:

$$V = nV_o^m e^{\frac{-p}{ap+B_o} + (\alpha_o + cp)T} \quad \text{(S-E-5)}, \tag{65}$$

and

$$T = \frac{\ln\left(\dfrac{V}{V_o}\right) + \dfrac{p}{B_o + ap}}{\alpha_o + cp} \quad \text{(S-E-5)}. \tag{66}$$

Using Eqs. (63) and (64) and substituting

$$f^n(p) = -\left(B_o + ap_{n-1}\right)\left[(\alpha_o + cp_{n-1})T - \ln\left(\frac{V}{V_o}\right)\right] \quad \text{(S-E-5)} \tag{67}$$

the pressure can be determined.

Investigating highly symmetrical atomic arrangements linear correlation between the volume coefficient of thermal expansion and the thermal heat capacity was detected[45]. Based on this correlation the integral

$$\int_{T=0}^{T} \alpha_{V_p} dT \tag{68}$$

is approximated by an area of trapezoid (Fig. 3) The integral below the Debye temperature $[T_\theta]$ is then

$$\int_{T=0}^{T \leq T_\theta} \alpha_{V_p}(T) dT \approx \left[\frac{\alpha_{V_p} T^2}{2T_\theta}\right]_{T \leq T_\theta} \tag{69}$$

while at temperatures higher than the Debye temperature is

$$\int_{T_\theta}^{T} \alpha_{V_p}(T) dT \approx \left[\alpha_{V_p}(T - T_\theta)\right]_{T_\theta}^{T > T_\theta}. \tag{70}$$

Combining the two parts Eqs. (69) and (70) gives the general formula

$$\int_{T=0}^{T} \alpha_{V_p}(T) \, dT \approx H(T) \frac{\alpha_{V_p} T^2}{2T_\theta} + [1 - H(T)] \, \alpha_{V_p}\left(T - \frac{T_\theta}{2}\right), \tag{71}$$

where H(T) is the Heaviside or unit step function defined as:

$$H(T) = \begin{cases} 0 & \text{if } T > T_\theta \\ 1 & \text{if } T \leq T_\theta \end{cases}. \tag{72}$$

Substituting Eq. (71) into Eq. (65) gives

$$V = nV_o^m e^{\frac{-p}{ap+B_o} + H(T)\frac{(cp+\alpha_o)T^2}{2T_\theta} + [1-H(T)] \, (cp+\alpha_o)\left(T - \frac{T_\theta}{2}\right)} \quad \text{(Debye)}. \tag{73}$$

$T_\theta = 1100K$ value was used for the calculations[46]. Assuming a linear pressure dependence for the Debye temperature requires the introduction of an additional multiplier g

$$T_\theta(p) = gp + T_\theta(p = 0). \tag{74}$$

Equation (73) can be written then as:

$$V = nV_o^m e^{\frac{-p}{ap+B_o} + H(T)\frac{(cp+\alpha_o)T^2}{2(gp+T_\theta)} + [1-H(T)](cp+\alpha_o)\left(T - \frac{gp+T_\theta}{2}\right)} \quad \text{(Debye + pressure)} \tag{75}$$

The validity of Eqs. (61), (65), (73), and (75) will be tested to experiments and I will label these equations as S-E-7, S-E-5, Debye, and Debye + pressure respectively.

## V. TESTING THE EoS TO EXPERIMENTS OF PEROVSKITE

Perovskite, the most abundant mineral of the mantle, has been extensively investigated at high pressures and temperatures. The availability of the wide pressure and temperature range experiments makes this mineral ideal for thermodynamic studies. Experiments up to 25-30 GPa pressure usually use multi-anvil apparatus while at higher pressures diamond anvil cells (DAC) are used. The experimental results of multi anvil press[47-51] and diamond anvil[52-54] are used in this study. The experiments covers the pressure 0-109 GPa, and the temperature 293-2199 K ranges. The distribution of the 269 experiments is shown on Fig. 4.

### A. Fitting criteria

The fitting accuracy of empirical EoSs with the same number of parameters is evaluated by correlation coefficients and RMSD. The fit quality of models using different numbers of parameters can not be evaluated by their correlation coefficients only[55-57]. The test devised assessing the right level of complexity is the Akaike Information Criteria AIC[58,59]. Assuming normally distributed errors, the criterion is calculated as:

$$AIC = 2k + n \ln\left(\frac{RSS}{n}\right), \tag{76}$$

where n is the number of observations, RSS is the residual sum of squares, and k is the number of parameters. The preferred model is the one which has the smallest AIC value.

### B. Fitting parameters

The fitting parameters, correlation coefficient, RMSD, and AIC were calculated for Eqs. (61), (65), (73), and (75). The best fit is achieved by Eq. (65). The trapezoid approximations used for the volume coefficient of thermal expansion in Eqs. (73) and (75) did not increase the fitting. The explanation could be that the introduced error in the low temperatures is small and overridden by the better fitting at higher temperatures. The parameters providing the best fit are:

$B_o = 267.5$ GPa, $V_o = 24.284$ cm$^3$, $\alpha_o = 2.079 \times 10^{-5}$ K$^{-1}$, $\frac{\partial B_o}{\partial p} = 1.556$ and

$\frac{\partial \alpha_o}{\partial p} = -1.098 \times 10^{-7}$ K$^{-1}$GPa$^{-1}$. Previous studies[1, 51 and refs. therein] reported 231-273 GPa for the bulk modulus of pure MgSiO$_3$. The calculated parameters of the EoS and the fitting parameters are given in Table 1.

Based on visual inspection the residuals seem to be random (Fig. 5). The RMSDs or uncertainties for the five parameters EoS are 0.043 cm$^3$, 0.79 GPa, and 125 K for the volume, pressure, and temperature respectively. These values are very close to the uncertainties of the experiments[60] indicating that the proposed EoS correctly describes the P-V-T relationship of perovskite.

Starting from 300 K the data set was separated into groups covering 200 K temperature range. Within this temperature range it was assumed that the condition is isothermal. For each temperature range the fitting parameters of the most widely used, finite strain, interatomic potential, and empirical isothermal EoSs Birch-Murnaghan [Eq. (17)], Vinet [Eq. (24)], and Roy-Roy [Eq. (26)] respectively were determined. The fitting parameters of the conventional bulk modulus [$B_T$ conv.] were also calculated

$$V = V_{0T} e^{\frac{-p}{ap+B_T}} \quad \text{(Conventional)}. \tag{77}$$

Using the parameters determined from the overall fitting the residuals to the experiments were determined. The RMSD and AIC values were calculated for each of the temperature ranges from the residuals.

The fitting of the S-E-5 was also calculated by using the average temperature of the experiments in each temperature region. The fitting parameters of the S-E-5 (p-V) are slightly weaker than the Birch-Murnaghan and Vinet p-V EoSs. The RMSD values of the S-E-5 (p-V) EoS are better than the RMDS values of the Roy-Roy EoS while the AIC values are better in three and weaker in three temperature ranges.

Using the overall parameters of the S-E-5 (p-V-T) EoS the RMSD and AIC values are superior in 6 temperature ranges while in 2 temperature ranges the conventional equations are better. If the S-E-5 (p-V-T) EoS is fitted specifically to the experiments of the given temperature range then the fitting parameters are superior in all temperature ranges. The results are given in Table 2.

## VI. CONCLUSIONS

It is suggested that the volume is the sum of the initial volume, the thermal volume at the given pressure and the elastic volume. Definitions describing the relationships of the temperature and pressure to these fundamental volume parts are defined. Using these definitions a semi-empirical seven parameter (p-V-T) EoS is derived. The semi-empirical EOS is tested to the available high pressure and temperature experiments of perovskite. Based on the fitting it is suggested that the temperature derivatives of the bulk modulus and the volume coefficient of thermal expansion are zero and that the pressure derivatives of these parameters are constant. The residuals of the EOS and experiments show random distribution for all of the variables. The uncertainties of the fitting are very close to the uncertainties of the experiments.

The most widely used isothermal EOSs, Birch-Murnaghan, Vinet, and Roy & Roy equations were compared to the semi-empirical EOS by separating the experiments into 200 K temperature ranges. Under pure isothermal conditions Birch-Murnaghan and Vinet EoS give a slightly better fit to the experiments than the semi-empirical p-V EoS. Based on the RMSD and AIC values the S-E-5 (p-V-T) EoS is superior in every temperature range to the conventional equations.

The additional advantages of the semi-empirical EoS are that the expression is simply and allows calculating any of the required variables.

**ACKNOWLEDGEMENT**

I would like to thank Alexandre Laugier for his encouragement and Mike Sukop for reading and commenting the manuscript. This research was supported by Florida International University Dissertation Year Fellowship.

TABLE I. P-V-T fitting parameters and results.

| 0-109 GPa [N=269] | $B_o$ [GPa] | $V_o$ [cm$^3$] | $\alpha_o$ [10$^{-5}$ K$^{-1}$] | a | b [10$^{-3}$] | c [10$^{-7}$] | d [10$^{-10}$] | g | R | RMSD | AIC |
|---|---|---|---|---|---|---|---|---|---|---|---|
| V(p,T) (S-E-7) | 272.5 | 24.287 | 1.961 | 1.384 | 1.427 | -1.081 | 6.140 | | 0.99960488 | 0.042 | -1687.2 |
| V(p,T) (S-E-6) | 271.5 | 24.278 | 2.075 | 1.426 | 1.078 | -1.120 | | | 0.99960360 | 0.042 | -1688.4 |
| V(p,T) (S-E-5) | 267.5 | 24.284 | 2.079 | 1.556 | | -1.098 | | | 0.99960089 | 0.043 | -1688.5 |
| V(p,T)(Debye) | 265.69 | 24.460 | 2.610 | 1.507 | | -1.510 | | | 0.99938700 | 0.053 | -1573.1 |
| V(p,T) (Debye+pres) | 264.56 | 24.463 | 2.489 | 1.487 | | -1.592 | | -6.499 | 0.99943865 | 0.051 | -1594.8 |
| p(V,T) (S-E-5) | | | | | | | | | | 0.792 | -125.6 |
| T(V,p) (S-E-5) | | | | | | | | | | 125.0 | 2597.8 |

TABLE II. P-V fitting parameters and results. The values in the parentheses are fixed. The numbers following the name of the EoS represents the number of parameters left open for the fitting.

| | a | $B_{0T}$ [GPa] | $V_{0T}$ [cm$^3$] | $B_{0T}'$ | $B_{0T}''$ | R | RMSD | AIC |
|---|---|---|---|---|---|---|---|---|
| T = 300-500 K | | | | | N = 86 | | | |
| V(p,T) (S-E-5-0) | | | | | | | 0.045 | -534.2 |
| V(p,T) (S-E-5) | | | | | | 0.9996721 | 0.043 | -530.3 |
| V(p) (S-E-5) (T=332K) | | | | | | 0.99948789 | 0.054 | -491.9 |
| V(p) ($K_T$ conv.; 3) | 1.441 | 268.2 | 24.457 | | | 0.99948789 | 0.054 | -495.9 |
| V(p) (Roy-Roy; 3) | | 267.1 | (24.473) | -0.537 | -0.00168 | 0.99947692 | 0.055 | -494.1 |
| p(V,T) (S-E-5-0 ) | | | | | | | 0.766 | -45.8 |
| p(V,T) (S-E-5 ) | | | | | | | 0.952 | 1.5 |
| p(V) (S-E-5 ) (T=332K) | | | | | | | 1.161 | 35.6 |
| p(V) (Birch; 3) | | 256.0 | 24.473 | 3.832 | | 0.99934394 | 1.059 | 15.9 |
| P(V) (Vinet; 3) | | 254.5 | 24.475 | 3.976 | | 0.99934135 | 1.061 | 16.2 |
| T(V, p) (S-E-5-0) | | | | | | | 120.8 | 824.6 |
| T(V, p) (S-E-5) | | | | | | | 61.4 | 726.4 |
| **T = 500-700 K** | | | | | **N = 38** | | | |
| V(p,T) (S-E-5-0 ) | | | | | | | 0.042 | -240.5 |
| V(p,T) (S-E-5) | | | | | | 0.99937901 | 0.038 | -238.1 |
| V(p) (S-E-5) | | | | | | 0.99918784 | 0.044 | -227.9 |
| V(p) (K$_T$ conv.; 3) | 1.433 | 271.7 | 24.571 | | | 0.99918784 | 0.044 | -231.9 |
| V(p) (Roy-Roy; 3) | | 264.4 | (24.604) | 0.0161 | -0.00375 | 0.99911565 | 0.046 | -228.7 |
| p(V,T) (S-E-5-0 ) | | | | | | | 0.999 | -0.1 |
| p(V,T) (S-E-5 ) | | | | | | | 0.534 | -37.7 |
| p(V) (S-E-5 ) (T=610K) | | | | | | | 0.624 | -25.8 |
| p(V) (Birch; 3) | | 256.1 | 24.604 | 3.890 | | 0.99951269 | 0.606 | -32.1 |
| P(V) (Vinet; 3) | | 254.8 | 24.607 | 4.035 | | 0.99951030 | 0.608 | -31.9 |
| T(V, p) (S-E-5-0) | | | | | | | 151.4 | 381.5 |
| T(V, p) (S-E-5) | | | | | | | 98.6 | 358.9 |

| | | | | | | | |
|---|---|---|---|---|---|---|---|
| T = 700-900 K | | | | | N = 42 | | |
| V(p,T) (S-E-5-0 ) | | | | | | 0.033 | -286.2 |
| V(p,T) (S-E-5) | | | | | 0.99965604 | 0.031 | -280.8 |
| V(p) (S-E-5) (T=789) | | | | | 0.99956274 | 0.035 | -270.8 |
| V(p) ($K_T$ conv.; 3) | 1.448 | 264.7 | 24.671 | | 0.99956274 | 0.035 | -274.8 |
| V(p) (Roy-Roy; 3) | | 257.4 | (24.707) | -0.740 | -0.00098 | 0.99948930 | 0.038 | -268.2 |
| p(V,T) (S-E-5-0 ) | | | | | | 0.929 | -6.2 |
| p(V,T) (S-E-5 ) | | | | | | 0.414 | -64.1 |
| p(V) (S-E-5 ) (T=789) | | | | | | 0.483 | -51.1 |
| p(V) (Birch; 3) | | 249.2 | 24.707 | 3.971 | | 0.99979650 | 0.449 | -61.2 |
| P(V) (Vinet; 3) | | 246.7 | 24.713 | 4.171 | | 0.99979472 | 0.451 | -60.9 |
| T(V, p) (S-E-5-0) | | | | | | 144.6 | 417.8 |
| T(V, p) (S-E-5) | | | | | | 100.2 | 397.0 |

| | | | | | | | | |
|---|---|---|---|---|---|---|---|---|
| **T = 900-1100 K** | | | | | **N = 28** | | | |
| V(p,T) (S-E-5-0 ) | | | | | | | 0.041 | -178.6 |
| V(p,T) (S-E-5) | | | | | | 0.99962291 | 0.040 | -170.7 |
| V(p) (S-E-5) (T=975) | | | | | | 0.99948823 | 0.046 | -162.1 |
| V(p) (K$_T$ conv.; 3) | 1.585 | 246.2 | 24.837 | | | 0.99948823 | 0.046 | -166.1 |
| V(p) (Roy-Roy; 3) | | 247.4 | (24.854) | 0.567 | -0.0062 | 0.99943899 | 0.048 | -163.5 |
| p(V,T) (S-E-5-0 ) | | | | | | | 0.545 | -34.0 |
| p(V,T) (S-E-5 ) | | | | | | | 0.601 | -18.5 |
| p(V) (S-E-5 ) (T=975) | | | | | | | 0.712 | -9.1 |
| p(V) (Birch; 3) | | 236.0 | 24.854 | 4.210 | | 0.99970397 | 0.658 | -17.4 |
| P(V) (Vinet; 3) | | 232.9 | 24.862 | 4.453 | | 0.99971131 | 0.650 | -18.1 |
| T(V, p) (S-E-5-0) | | | | | | | 89.9 | 251.9 |
| T(V, p) (S-E-5) | | | | | | | 136.8 | 275.6 |

| T = 1100-1300 K | | | | | N = 26 | | |
| --- | --- | --- | --- | --- | --- | --- | --- |
| V(p,T) (S-E-5-0 ) | | | | | | 0.034 | -175.6 |
| V(p,T) (S-E-5) | | | | | 0.99982864 | 0.030 | -173.1 |
| V(p) (S-E-5) (T=1179K) | | | | | 0.99966371 | 0.041 | -155.6 |
| V(p) (K$_T$ conv.; 3) | 1.589 | 236.9 | 25.015 | | 0.99964480 | 0.044 | -138.1 |
| V(p) (Roy-Roy; 3) | | 243.9 | (25.006) | 0.649 | -0.0066 | 0.99956543 | 0.048 | -133.5 |
| p(V,T) (S-E-5-0 ) | | | | | | 0.794 | -12.0 |
| p(V,T) (S-E-5 ) | | | | | | 0.636 | -13.3 |
| p(V) (S-E-5 ) (T=1179K) | | | | | | 0.772 | -3.5 |
| p(V) (Birch; 3) | | 229.3 | 25.006 | 4.264 | 0.99973450 | 0.775 | -5.7 |
| P(V) (Vinet; 3) | | 223.5 | 25.030 | 4.589 | 0.99974031 | 0.766 | -6.2 |
| T(V, p) (S-E-5-0) | | | | | | 131.7 | 253.8 |
| T(V, p) (S-E-5) | | | | | | 137.4 | 266.0 |
| **T = 1300-1500 K** | | | | | **N = 17** | | |
| V(p,T) (S-E-5-0 ) | | | | | | 0.041 | -108.2 |
| V(p,T) (S-E-5) | | | | | 0.99972910 | 0.038 | -101.4 |
| V(p) (S-E-5) (T=1362K) | | | | | 0.99969173 | 0.040 | -99.2 |
| V(p) ($K_T$ conv.; 3) | 1.379 | 267.5 | 24.899 | | 0.99971664 | 0.040 | -84.3 |
| V(p) (Roy-Roy; 3) | | 250.8 | (25.039) | 0.641 | -0.00642 | 0.99963913 | 0.045 | -81.0 |
| p(V,T) (S-E-5-0 ) | | | | | | 0.694 | -12.4 |
| p(V,T) (S-E-5 ) | | | | | | 0.627 | -5.9 |
| p(V) (S-E-5 ) (T=1362K) | | | | | | 0.762 | 0.8 |
| p(V) (Birch; 3) | | 235.9 | 25.039 | 4.094 | | 0.99982165 | 0.678 | -4.9 |
| P(V) (Vinet; 3) | | 227.5 | 25.081 | 4.456 | | 0.99982355 | 0.675 | -5.0 |
| T(V, p) (S-E-5-0) | | | | | | 113.9 | 161.0 |
| T(V, p) (S-E-5) | | | | | | 120.4 | 172.9 |

| | | | | | | | | |
|---|---|---|---|---|---|---|---|---|
| **T = 1500-1700 K** | | | | | **N = 13** | | | |
| V(p,T) (S-E-5-0 ) | | | | | | | 0.041 | -83.2 |
| V(p) (K$_T$ conv.; 3) | 2.064 | 115.1 | 26.768 | | | 0.99949113 | 0.037 | -60.1 |
| V(p) (Roy-Roy; 3) | | 240.3 | (25.010) | 3.932 | -0.0205 | 0.99972832 | 0.046 | -55.6 |
| p(V,T) (S-E-5-0) | | | | | | | 0.468 | -19.7 |
| p(V) (Birch; 3) | | 240.6 | 25.010 | (4.0) | | 0.99982114 | 0.804 | -0.4 |
| P(V) (Vinet; 3) | | 247.2 | 24.962 | (4.0) | | 0.99981647 | 0.814 | -0.1 |
| T(V, p) (S-E-5-0) | | | | | | | 77.2 | 113.0 |
| **T = 1700-1900 K** | | | | | **N = 13** | | | |
| V(p,T) (S-E-5-0 ) | | | | | | | 0.058 | -74.2 |
| V(p) (K$_T$ conv.; 3) | 2.341 | 113.3 | 26.806 | | | 0.99582439 | 0.057 | -51.2 |
| V(p) (Roy-Roy; 3) | | 218.7 | (25.538) | 0.527 | -0.00680 | 0.99554394 | 0.059 | -50.6 |
| p(V,T) (S-E-5-0 ) | | | | | | | 0.630 | -9.0 |
| p(V) (Birch; 3) | | 210.1 | 25.538 | (4.0) | | 0.99633845 | 0.837 | 0.4 |
| P(V) (Vinet; 3) | | 213.6 | 25.509 | (4.0) | | 0.99627391 | 0.844 | 0.6 |
| T(V, p) (S-E-5-0) | | | | | | | 117.0 | 123.8 |

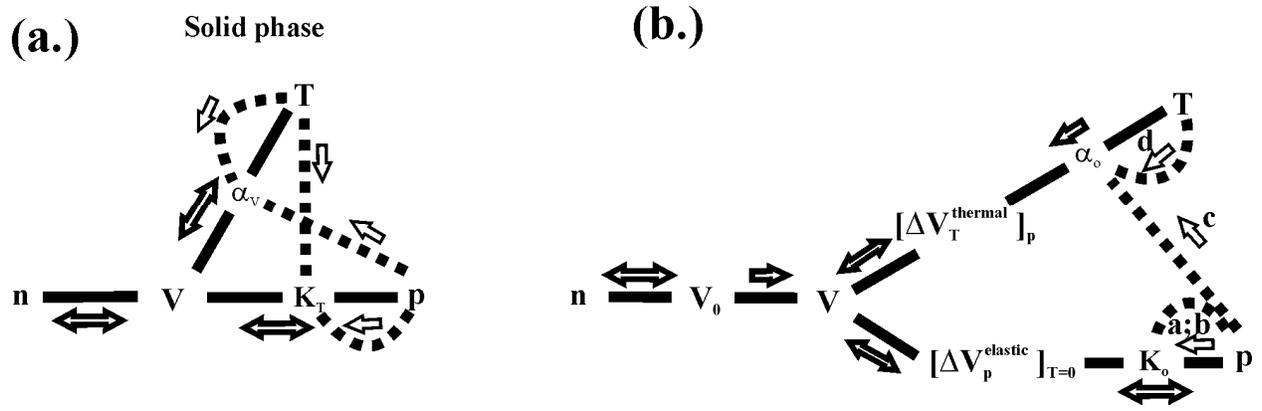

FIG. 1. Thermo-physical relationships (a) solid phase conventional description (b) proposed description (The arrow ↔ represent a reversible while → represents an irreversible relationship or process.)

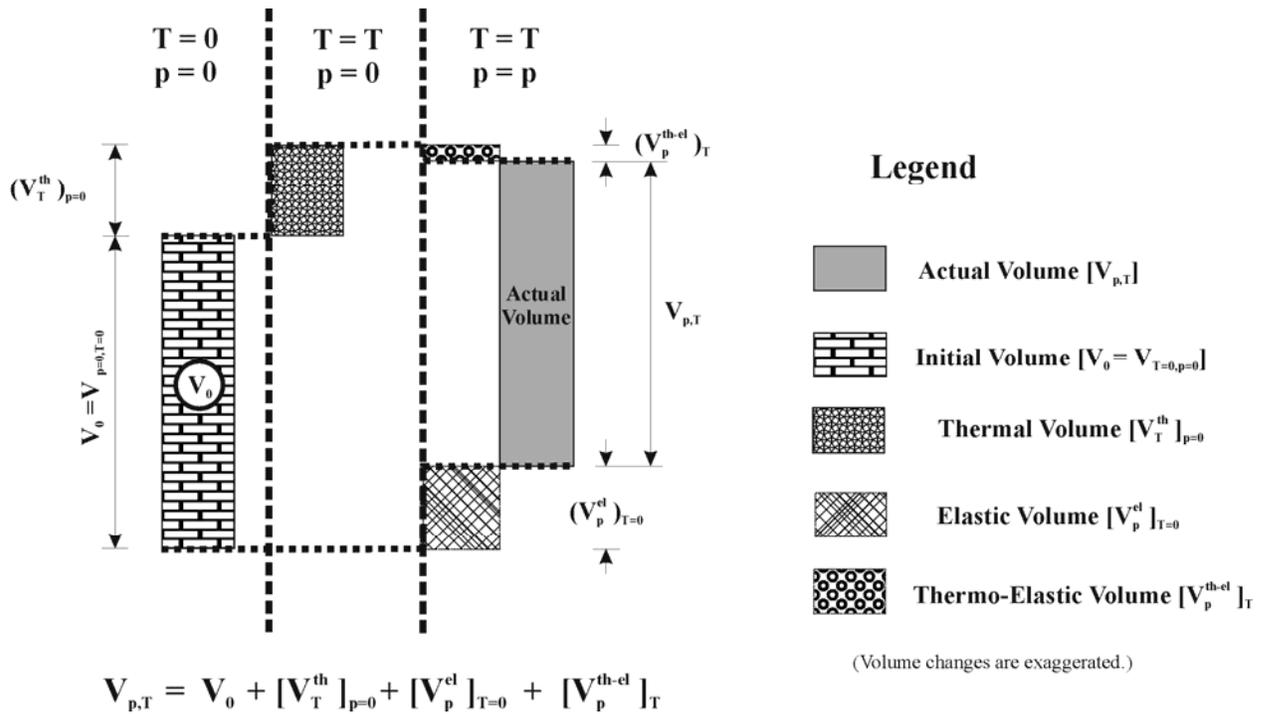

FIG. 2. The fundamental volume components of the actual volume.

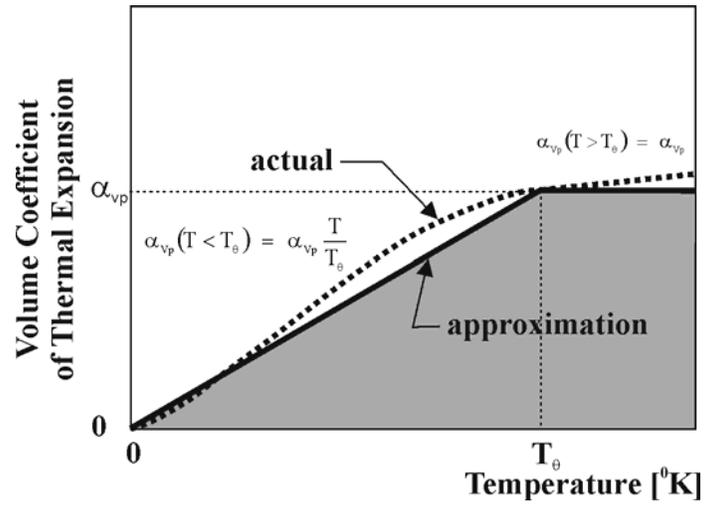

FIG. 3. Approximation used for the volume coefficient of thermal expansion in Eqs. (73), and (75).

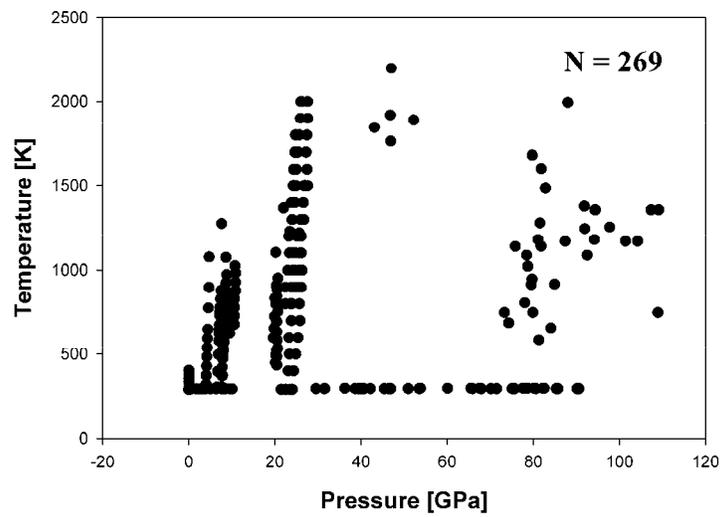

FIG. 4. Pressure-temperature range covered by the experiments of perovskite.

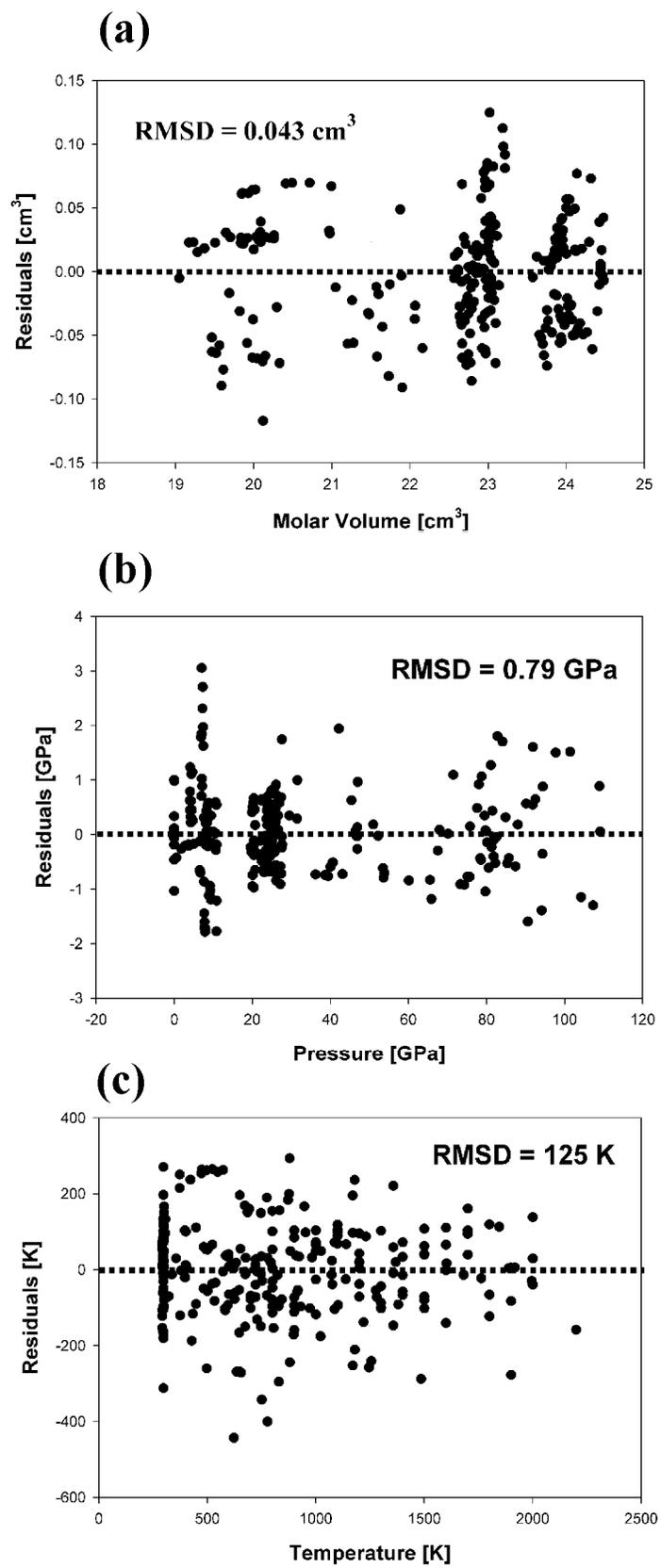

FIG. 5. The residuals plotted against (a) volume (b) pressure (c) temperature.